# Tube-Tube and Tube-Surface Interactions in Straight Suspended Carbon Nanotube Structures

Z. R. Abrams and Y. Hanein*

*School of Electrical Engineering, Department of Physical Electronics, Tel-Aviv University, Tel-Aviv 69978, Israel*

* Corresponding author, Phone: 972-3-6407698, Email: hanein@eng.tau.ac.il

An investigation concerning the tautness of suspended carbon nanotubes (CNTs), grown using the chemical vapor deposition (CVD) method is presented. The suspended nanotubes were analyzed with both a transmission electron microscope (TEM), and a high resolution scanning electron microscope (HR-SEM). The HR-SEM and TEM investigations revealed that the interactions between CNTs among themselves, and with the surface on which they are grown, is a primary cause for the tautness of suspended tubes. Specifically, the tube-tube and tube-surface dynamics cause adjoining tubes to create a "zipper effect", thereby straightening and tightening them. Suspended CNTs cling to each other, and to as much of the surface as possible, in order to minimize their total energy, creating taut, suspended structures. This effect can be so strong so as to force wide tubes to buckle, with no other external force involved. The implications of this study include all forms of alignment processes of nanotubes using the CVD method. The results presented here provide the groundwork for the capability of fine-tuning the control of CNT network formation using substrate mechanical features.

Carbon nanotubes are quickly becoming one of the most widely used novel materials for nanotechnology[1,2] and nanofabrication. Individual CNTs, comprised of a single rolled-up layer of graphite, and known as single walled carbon nanotubes (SWNTs), are specifically the impetus for a plethora of new research due to their unique and simple structure. SWNTs have been shown to be viable as nano-components for logic circuits[3], field-effect transistors[4], and optical detectors[5], to name a few. However, one major drawback inherent to these structures is their fabrication, which provides little control over both their exact structure and placement. This paper examines CNTs grown by CVD, which governs the placement of CNTs on a substrate.

Using the CVD method, CNTs have been shown to grow in relative isolation and alignment using a number of methods[6,7,8,9,10,11]. Particularly, many reports of aligned CNTs have used either trenches in silicon, or silicon pillars to isolate CNTs and/or align them[6,7,10,11]. In many of these studies, the resulting suspended CNTs are seen as being tightly stretched between either the pillars or the trenches, yet, as beneficial as this isolation and alignment method may be, the effect involved has been little understood. In an effort to reveal the origins of the tautness in these structures, we grew CNTs between silicon pillars, and on TEM grids containing large holes, and analyzed them using both a HR-SEM, and a TEM. Both pillars and TEM grids were found to adequately model the bridging-nanotube growth. The findings presented here indicate that some CNTs bridging gaps are, in fact, pairs or bundles, while others are single-bridging nanotubes, in which the surface to tube interaction has a decisive role in their straightness.

A brief description of the fabrication of the pillars and the CVD process employed follows. Pillars were created by patterning an array of circles, using standard photolithography, on a silicon dioxide surface (450μm Si, 500nm $SiO_2$). After developing the resist, the $SiO_2$ was etched using reactive ion etch, followed by a deep-reactive ion etch of the silicon substrate, forming pillars with heights of 17μm. Spacing between pillars arrayed in a grid was 16, 20 and 25μm. The CVD growth of CNTs utilized the following protocol: Substrates were coated with a $Fe(NO_3)_2$ catalyst, which was suspended in isopropyl alcohol (IPA), after having been sonicated, and centrifuged[12]. The substrates were placed in a tube furnace, and brought to 900°C under a constant flow of hydrogen gas, then, ethylene (C$_2$H$_4$) was introduced for 9 min[13], after which the substrates were allowed to cool down to room temperature. TEM grids (DuraSiN mesh) were prepared for growth in a similar fashion to the silicon pillars. The meshes used comprised of either 40μm square holes, or 2μm circular holes, in a silicon nitride thin-film, supported by a silicon substrate, and the viewing area of the grid was a 500×500μm window.

The limiting resolution of the HR-SEM prevents one from differentiating between two adjoining tubes, and the silicon substrates are not compatible with viewing in a TEM[14]. Therefore, TEM grids were used to verify the true nature and origin of the suspended nanotubes' tautness: specifically, to discover whether some of the suspended tubes were individual SWNT, as previously posited, or "bundles" of two or more tubes, and to ascertain the cause of the individually isolated SWNTs tautness. The choice of 40μm holes in the nitride was found to be sufficiently equivalent to both pillar to pillar and trench growth. CNTs were seen to grow over the mesh holes in a similar fashion to the growth between pillars (see Fig.1(c)).

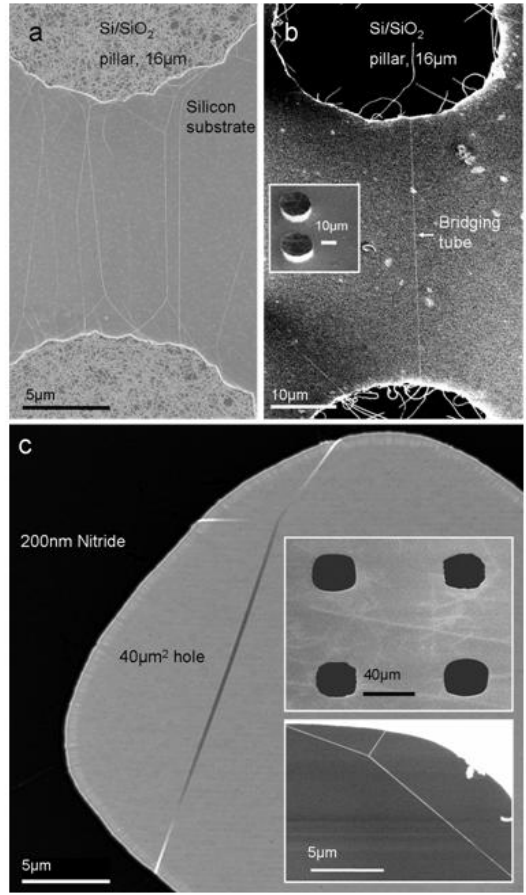

Figure 1: (a) Top-down HR-SEM image of oxide capped silicon pillars, with many taut, suspended bridging nanotubes, many of which end in y-junctions. The large number of bridging tubes is due to the high density of catalyst used. (b) A single, taut, bridging tube. Inset: an isometric view of the 17µm high pillars. (c) A TEM image of a suspended nanotube bridging a 40µm hole in a 200nm silicon nitride thin film. The nitride film is opaque in the TEM. The Y-shaped set of tightly suspended tubes can easily be seen at this magnification (see text). Upper inset: an HR-SEM image of the TEM grid, CNTs can be seen on the surface around the holes (thin white lines). Lower inset: an HR-SEM image of the same tube.

Figure 1 shows typical results of a CVD growth, with CNTs growing from the catalyst, which is adsorbed onto the surface of the pillars/grid. As can be seen in Figs. 1 (a) and (b), CNTs bridge the gap between adjacent pillars (white lines). This growth is spontaneous, and the prevalence of nearest-neighbor attachment of CNTs between pillars could be found throughout the array of pillars. The primary difference between Figs. 1(a) and (b) is the density of catalyst used. In Fig. 1(a), with a higher density of catalyst deposited, many bridging nanotubes are seen, some of which end in y-junctions adjacent to the pillar edge. Few of these suspended CNTs are y-junction free, which is similar to the situation depicted in figure 1(b). The results of Fig. 1(b) are also similar to those in previous studies concerning bridging nanotubes. In almost all cases of bridging CNTs, the suspended tubes are visibly taut, and straight, however the y-junction can only be seen in some of the suspended tubes. Figure 1(c) displays a typical 40μm hole in a TEM mesh, with a y-junction made of suspended CNTs. The spontaneous growth of the CNTs over the holes using the CVD process can therefore be effectively described as akin to the results of Figs. 1(a) and (b). A brief side-note must be made regarding the image in Fig.1(c): Although the scale bar is 5μm, the diameters of the nanotubes were around 2-4nm. The distortion due to extreme Fresnel fringes associated with the nearly one-dimensional CNTs made them visible even at such low magnification[15].

It has been previously postulated[6,16] that bridging nanotube growth is due to the CNTs growing horizontally from the pillar surface tops, and swinging in the air due to thermal vibrations, whereupon they stick to neighboring pillars upon contact. Nanotubes were modeled as vibrating cantilevers, wavering as much as 6μm at a length of 20μm. However, this model does not account for the discernible tautness seen in either the structures used here, or in other similar structures, and a better description is desirable.

A better understanding of the process taking place can be achieved through careful inspection of the TEM and HR-SEM results. This enabled us to verify and study the distinction between the two types of suspended CNT growth, namely nanotubes ending in y-junctions, and others with no apparent y-junctions. As will be elaborated presently, the former will be described in terms of tube-tube interactions, and the latter in terms of tube-surface interactions.

The first case is relatively easy to explain by considering the interactions between two bridging nanotubes (Fig. 2). Two coextending tubes, vacillating in the thermal field permeating the CVD growth, will occasionally come into contact. The pair will then adhere to each other through the van der Waals (vdW) force attracting the chains of carbon atoms. The resultant effect is similar to a zipper action between the two anchor points (Fig. 2(a)). This model is verified by the close-up images of the y-junctions in the TEM, which clearly reveal the nature of the y-junction as being a juxtaposition of at least two nanotubes (Figs. 2(b), (c)). In some cases, the nanotubes will also buckle. Figure 2(b) shows a TEM image of such a case. Figure 2(c) shows the nanotubes merging at a shallow angle, with the rightmost pair as being warped. Figure 2(d) shows a typical HR-SEM close-up of a y-junction. The bottom CNT, extending to an opposite pillar, is distinctly visible as being a bundled pair, as it is a combination of two y-junctions. Here as well, the difference between buckled and warped tubes is apparent. A more detailed description of this effect will follow.

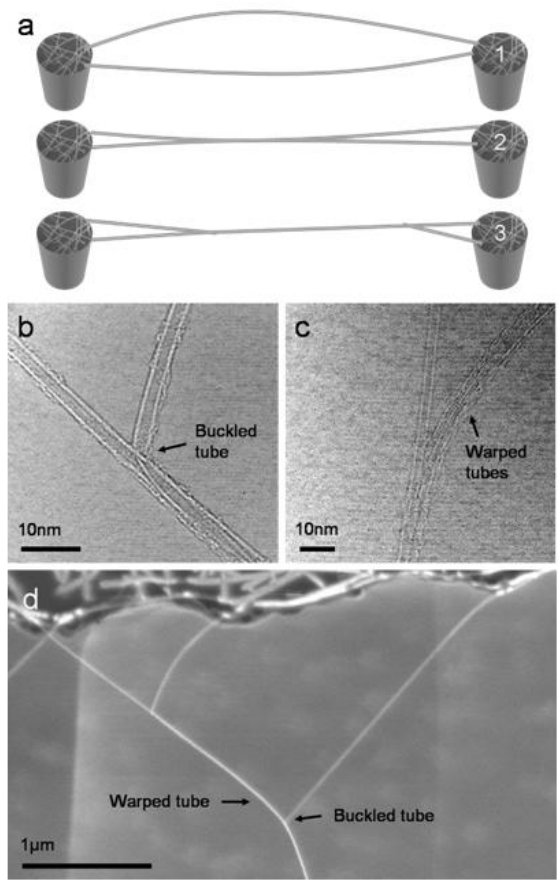

Figure 2: (a) Schematic of the "zipper effect" between a pair of coextending nanotubes. The final result is a set of y-junctions between the pillars. (b) TEM image of a set of adjoining tubes, with the rightmost tube buckling, and (c), a set of CNTs converging at a low angle. (d) HR-SEM close-up of a cascading y-junction. The leftmost set of CNTs is warped, whereas the rightmost one is apparently buckled, as seen by the angle of contact.

The second effect mentioned above, in which there are no y-junctions, can be explained in terms of tube-surface interactions (Fig. 3). In Fig. 3(a), this effect is illustrated as an extended tube reaching a taut equilibrium point by allowing its slack segments to adhere to the surface edge of the pillars. The tube-surface interaction, which is also derived from the vdW forces between them, creates a similar effect to the previously described zipper-effect, tightening and straightening the suspended nanotube.

Figures 3(b) and (c) demonstrate the contour following of the tube segments to the surface, due to this interaction. Fig. 3(b) shows a TEM example of how much this interaction can bend/buckle a tube. The single SWNT in this image extended over a 2μm diameter hole in the grid, and, because of the tube-surface adherence appearing in the image, was relatively taut. This effect is manifest in many of the HR-SEM images as well, and in Fig. 3(c), an extending SWNT is shown bending along the edge of a pillar top. The tube-surface adherence can be seen to follow the contours of the pillar edge over the course of almost a micron (several samples displayed contour following for several microns).

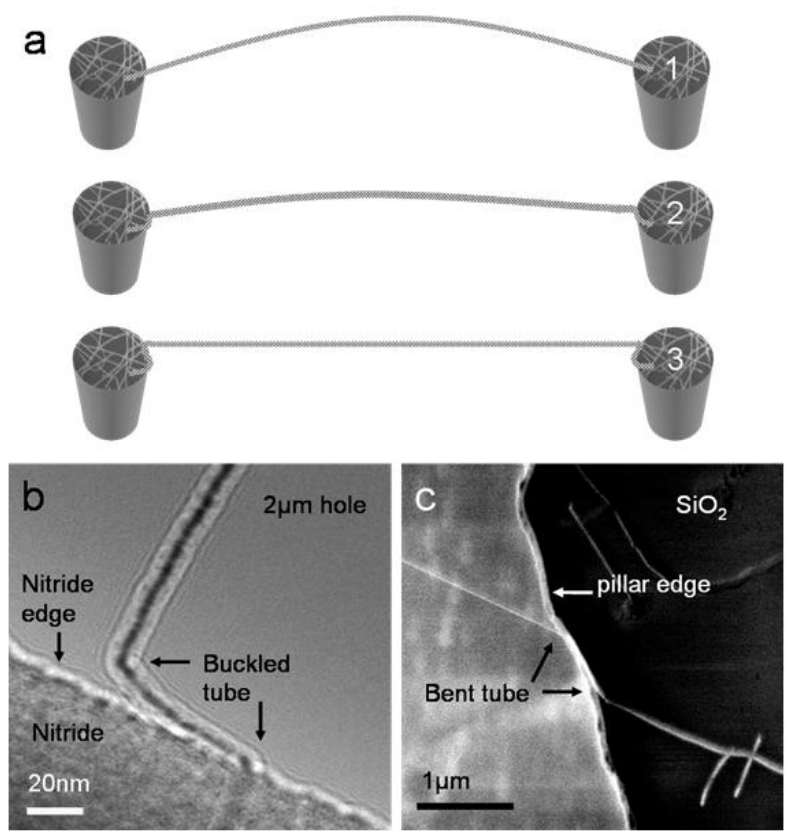

Figure 3: (a) Schematic of the effects of the tube-surface interactions in creating a taut SWNT bridging the gap between two pillars. (b) TEM image of a SWNT following the contour of a 2μm hole, creating a taut SWNT bridging the hole. (c) HR-SEM image of an extending SWNT adhering to as much of the pillar edge as possible, over a scale of almost a micron. The resolution of the HR-SEM limits the capability of affirming either the warping, or buckling of the nanotube.

To better understand the results described above, we now turn our attention to the forces involved. A theory regarding the binding energy between parallel tubes has been previously given[17,18]. There, a single, free-standing, pair of tubes was analyzed (i.e., only one end was anchored). The analysis took into consideration the bending stiffness of a CNT, when taken as an infinite beam, given by:

$$E \cdot I = E \cdot \frac{\pi}{64}[d^4 - (d-2t)^4] \quad (1)$$

where $EI$ is the bending stiffness, and $I$, $E$, $d$, and $t$ are the moment of inertia, Young's modulus, outer diameter, and CNT wall thickness, respectively. In our investigation, most of the CNTs viewed in the TEM were SWNT, however, some multilayered CNTs were seen as well. The Young's modulus of a CNT is in the range of ~1TPa[19] and the diameters of the CNTs measured ranged from 1.7-5.8nm. The model takes into account the elastic and binding energies of a pair of adjoining nanotubes (Fig 2(c)). The total energy, $F$, given in ref. [17] as:

$$F = \frac{4EI}{l}\left(\theta^2 - 3\theta\frac{\Delta}{l} + 3\frac{\Delta^2}{l^2}\right) + \gamma l \quad (2)$$

with $\theta$, $\Delta$, $l$, $\gamma$ as the angle extended between the point of contact and the separated tubes, distance between the separated tubes, distance from the binding point, and the binding energy between the tubes, respectively. Equation 2 represents the interplay between the elastic energy of the tube (i.e., warping), and the binding energy, whereas the system aims at reaching a minimum.

As seen in Fig. 2(b), CNTs are capable of buckling in the same way a plastic straw would, when a critical value of stress is induced. In this case, the system does not reach the equilibrium defined by Eq. 2, rather, it will reach a point of equilibrium defined by the strain after the buckling point, which becomes linear with the stress induced[20]. By viewing several y-junctions from the TEM images, it was found that cases of non-buckling, as in Fig. 2(c) consisted of nanotubes with diameters of 1.7-3.4nm, whereas the buckled sets had diameters of 4-5.9nm. This is consistent with the calculated reports of the critical stress of CNTs[21], which is inversely proportional to the diameter as: $\sigma_{crit} \propto 1/r$. Larger diameter tubes therefore have a higher tendency of buckling under the stress of the zipper-effect, both in the case of tube-tube, and tube-surface interactions. Furthermore, a bundle of tubes would be more resilient to buckling, as is demonstrated in Fig. 2(d).

The preceding results may be summarized as follows: A slack, suspended CNT will almost always become straight and taut due to either of the two vdW zipper-effects described above: tube-tube and tube-surface interactions. A combination of the two is also possible. Figure 4 demonstrates a combination of the two interactions described here. Figure 4(a) shows a single SWNT attached to both the edge of a 40μm hole in a nitride mesh, as well as its adherence to a pair of suspended CNTs that bridged a large section of the hole. The segments are visibly taut, as a result of the interactions involved. As seen in the close-up in Fig. 4(b), the single CNT undergoes a strong buckling effect, merely due to the presence of the tube-tube interaction. The system represented in Fig. 4(a) can be seen as the minimal-energy distribution possible under this configuration. Deformations of CNTs resulting in the buckling of a tube are readily found in the setup involving the growth of CNTs on TEM meshes, and buckling at angles ranging from 53-150° were seen[22], as is evident in Figs. 2(b) and 4(b), including the inset.

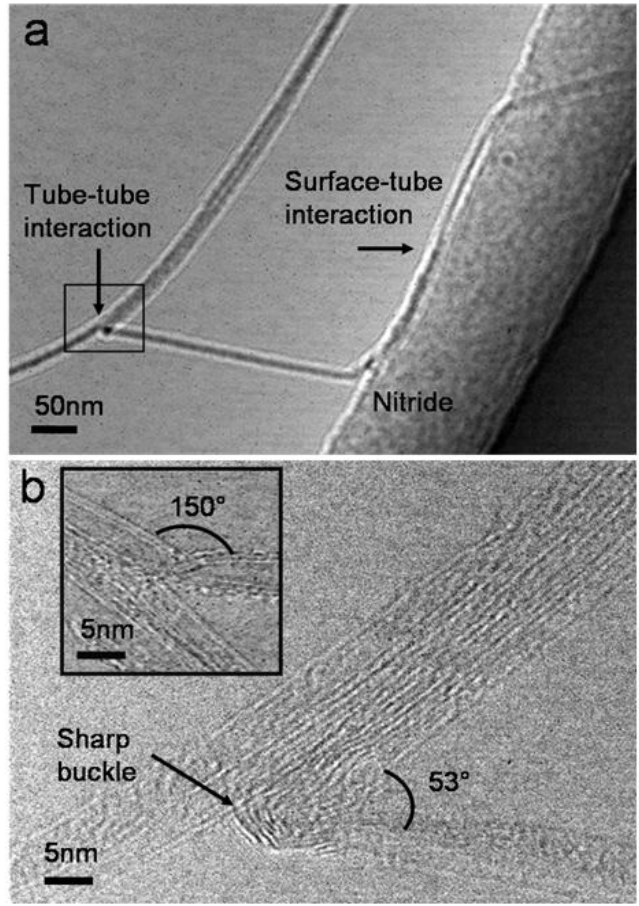

Figure 4: (a) TEM of a combination of both of the interactions described in the text. A pair of CNTs suspended over a 40μm hole in a nitride mesh is seen attached to a single nanotube, which is taut between the nitride edge and the point of contact. The tube-surface interaction is notably seen as playing a role. (b) A close up of the junction in (a), in which an extreme buckling of the CNT is seen. Inset: a more conventional example of CNT buckling.

To conclude, the growth of suspended CNTs is seen to be highly dependant on the interactions of the materials used, whether in the case of nanotube to nanotube vdW interaction, or nanotube to oxide/nitride interaction. The zipper-effect is seen to be strong enough to cause wider-diameter SWNTs to buckle under the adhering forces involved and required no external forces[20,23]. While many studies including simulation of CNT buckling have been previously described[24], the growth method upon large nitride holes presents a novel platform for further empirical studies of this phenomenon, as well as other mechanical properties of CNTs. Nevertheless, one of the downsides of these TEM grids is that when imaging long, taut tubes for extended periods of time, the nanotubes tended to come apart. This is probably due to the lack of heat dissipation along the nanotube[25], caused by the TEM's beam. Furthermore, a more detailed analysis of the suspended tubes, which would include the stretching of the tubes, is prudent, especially in light of recent evidence of extreme elongation of CNTs[26]. Finally, an empirical study of the critical stress of SWNTs can be implemented using this setup, and will be studied in the future.

The immediate implications of the results in this letter are twofold: The first is that some long, suspended nanotubes are in fact bundles, or pairs. The second is that the effect of the tube-surface interaction in all forms of CVD CNT growth, whether suspended or not, perceptibly affects the directionality of nanotubes on various surfaces. While the negative implications of the first conclusion would seem to be that many bridging tubes are *not* individual SWNT, seemingly diminishing their viability as electronic, or electro-optic devices, it is important to note that adjacent nanotubes have a strong tunneling effect between them[27]. Moreover, the major ramification of the second conclusion is the verification of the possibility of creating individually isolated SWNT, in a straight, taut network. The possibilities inherent in the methods utilized by either growing CNTs over trenches or between pillars lies in their ability to create viable CNT networks. In either case, the straightness of these suspended tubes is advantageous, as it limits the number of crystal defects in the pair, as opposed to those grown on the substrate[28], especially in lieu of the evidence presented here regarding tube-surface interactions. By fine-tuning the process, especially in terms of catalyst density and placement, an organized network of long, taut, individually isolated SWNTs can be fully realized by utilizing the effects noted in this letter.

Acknowledgments: The authors thank Dr. Yossi Lereah and Dr. Alexander Tsukernik for their help with the TEM and HR-SEM, respectively. This research was supported in part by an ISF grant.

[15] In a 200keV TEM beam, the diameters of the CNTs were verified at ×125,000 magnification, yet, the distortion in the image makes them appear much thicker. An additional method of acquiring the same images would be to bring the substrate into extreme de-focus. For example, when the surface of the substrate was brought to zero z-axis focus, the lens needed to be defocused by 260μm, which was exactly the distance the substrate was shifted in order to bring it into eccentric focus. However, this method is less preferable, as it requires the continual change of the intensity of the beam.

[16] Y. Homma, Y. Kobayashi, T. Ogino, T. Yamashitab, *Appl. Phys. Lett.*, 81, 2261 (2002)

[17] B. Chen, M. Gao, J. M. Zuo, S. Qu, B. Liu, Y. Huang, *Appl. Phys Lett.*, **83**, 3570 (2003)

[18] L. A. Girifalco, M. Hodak, R.S. Lee, *Phys. Rev. B*, **62**, 13104 (2000)

[19] M. F. Yu, B. S. Files, S. Arepalli, R. S. Ruoff, *Phys. Rev. Lett,.* **84**, 5552 (2000)

[20] S. Iijima, C. Brabec, A. Maiti, J. Bernholc, *J. Chem. Phys*., **104**, 2089 (1996)

[21] Y. Wang, X. X. Wang, X. G. Ni, H. A. Wu, *Comp. Mater. Sci*, **32**, 141 (2005)

[22] With the angle as defined in figure 4(b).

[23] J. F. Despres, E. Daguerre, K. Lafdi, *Carbon*, **33**, 87 (1995)

[24] K. M. Liew, C. H. Wong, M. J. Tan, *Appl. Phys. Lett.*, **87**, 041901 (2005)

[25] M. Kuroda, A. Cangellaris, J. P. Leburton, *Phys. Rev. Lett.*, **95**, 266803 (2005)

[26] J. Y. Huang, S. Chen, Z. Q. Wang, K. Kempa, Y. M. Wang, S. H. Jo, G. Chen, M. S. Dresselhaus, Z. F. Ren, *Nature*, **439**, 281 (2006)

[27] H. Stahl, J. Appenzeller, R. Martel, P. Avouris, B. Lengeler, *Phys. Rev. Lett.*, **85**, 24 (2000)

[28] T. Hertel, R. E. Walkup, P. Avouris, *Physical Review B*, **58**, 13870 (1998)